\documentclass[sigconf]{acmart}
\AtBeginDocument{%
  }

\acmConference[EASE 2026]{The 30th International Conference on Evaluation and Assessment in Software Engineering}{9--12 June, 2026}{Glasgow, Scotland, United Kingdom}

\usepackage{graphicx}
\usepackage{textcomp}
\usepackage{xcolor}
\usepackage{diagbox}
\usepackage{multirow}
\usepackage{makecell}
\usepackage{tikz}
\usepackage{tabularx}
\usetikzlibrary{shadows, shapes.misc}
\usepackage{booktabs}

\begin{document}

\settopmatter{printacmref=false}
\setcopyright{none}
\renewcommand\footnotetextcopyrightpermission[1]{}
\pagestyle{plain}

\title{Tail-aware N-version Machine Learning Models for Reliable API Recommendation}

\author{Aoi Matsuda}
\affiliation{%
  \institution{University of Tsukuba}
  \city{Tsukuba}
  \country{Japan}
}
\email{matsuda.aoi@sd.cs.tsukuba.ac.jp}

\author{Fumio Machida}
\affiliation{%
  \institution{University of Tsukuba}
  \city{Tsukuba}
  \country{Japan}
}
\email{machida@cs.tsukuba.ac.jp}

\author{David Lo}
\affiliation{%
  \institution{Singapore Management University}
  \country{Singapore}
}
\email{davidlo@smu.edu.sg}

\begin{abstract}
Machine learning (ML)-based API recommendation helps developers efficiently identify suitable APIs to complement the application code. However, code datasets used to train ML models often exhibit a long-tail distribution, leading to unreliable API recommendations, especially for infrequently used API methods at the tail of the distribution. 
To address this issue, we propose N-version API Recommendation (NvRec), which leverages N different versions of ML models to enhance the reliability of API sequence recommendations by suppressing unreliable outputs entailing tail APIs. NvRec leverages a set of available ML models and profiles their performance on individual API methods with their tail properties. The generated model profile is used at inference time to filter out unreliable API recommendations and determine the final output.
We implement NvRec using five API recommendation models, including CodeBERT, CodeT5, MulaRec, UniXcoder, and CodeT5+, and evaluate it on a public benchmark dataset constructed from compilable Java projects. 
For the three-version NvRec, we find that the combination of CodeT5, MulaRec, and UniXcoder achieves the highest true accept rate of 83.8\%, with a rejection rate of 80.7\%, when majority voting is restricted to highly reliable candidates. In contrast, the five-version configuration achieves its highest true accept rate of 83.1\% with simple majority voting, while reducing the rejection rate to 69.0\%. Overall, the five-version configuration offers a better balance between true accept rate and rejection rate.

\end{abstract}


\maketitle
\section{Introduction}
Application Programming Interfaces (APIs) define the methods for accessing the functionality provided by software libraries and frameworks. Software developers call library functions through the API to efficiently implement the application code. 
However, to identify appropriate methods in a given application context, developers should comprehend the functionality of individual APIs from their documentation \cite{APIChange, ImprovingAPIDoc}, which is prohibitively demanding when there are a vast number of available APIs. 
Although many coding tasks can be automated by AI and coding agents \cite{MetaGPT, AgentCoder}, human developers still need to be responsible for the code to ensure reliability.

To support developers in finding appropriate API methods, several recommendation techniques have been presented. The conventional API recommendation techniques are largely divided into single-API recommendations \cite{Identifier-based} and API method sequence recommendations \cite{DeepAPI, MULAREC}. 
For example, Heinemann et al. \cite{Identifier-based} proposed a single API recommendation technique that analyzes identifiers, such as variable names and method names, in the source code and recommends a relevant API method based on the code's context. 
For the API method sequence recommendation, Gu et al. \cite{DeepAPI} proposed DeepAPI, which uses a deep learning model to generate API method sequences from natural language queries. Recently, Large Language Models (LLMs) have also been used for API recommendation \cite{MULAREC, DDASR, Gorilla, Toolformer}. For instance, Irsan et al. \cite{MULAREC} achieved higher-precision API method sequence recommendation by inputting both natural language queries and source code into an ML model. 
Ma et al. presented CAPIR \cite{CAPIR}, which decomposes coarse-grained requirements into fine-grained ones and composes API method sequences by retrieving relevant APIs with an LLM.

Although machine learning (ML)-based API recommendations achieve remarkable performance, ML models inherently depend on large source code datasets for training. The distribution of datasets for programming tasks often exhibits a long-tail property, which poses a challenge to the quality of API recommendations, especially for tail API methods \cite{TheDevil}. Recommendation results containing tail API methods tend to be unreliable. Adopting such unreliable API recommendations can introduce bugs into the resulting software, leading to additional development and debugging costs.

To address the gap, this paper proposes the N-version API method sequence Recommendation (NvRec) to provide reliable API method sequence recommendations by leveraging the diversity of multi-version ML models to filter out unreliable outputs that contain tail API methods.
Inspired by N-version programming \cite{avizienis1985n} and the N-version ML system architecture \cite{machida2023using}, NvRec uses multiple versions of ML models to produce reliable outputs by aggregating their individual inference results.
Since each ML model has a different predictive capability, we prepare the \textit{model profile}, which records the performance of individual ML models and the tail property for each API method. At inference time, we introduce \textit{Tail Analyzer} to check if the inputs are categorized as a tail case where an API recommendation is likely to have tail API methods. If the input is classified as a tail case, the request is dropped to suppress unreliable recommendations. If the request passes \textit{Tail Analyzer}, API method sequence candidates are generated by N-version ML models (multi-version inference). Next, unreliable candidates are excluded using the model profile (filtering). Finally, the recommendation result is selected from the remaining candidates by a specific output decision method (output decision). By leveraging the diversity of ML models and filtering out unreliable API recommendations with tail APIs, NvRec can recommend only reliable API method sequences. 

We implemented NvRec using several ML models for API recommendation tasks. CodeBERT \cite{CodeBERT}, CodeT5 \cite{CodeT5}, and MulaRec \cite{MULAREC} are chosen from the previous studies, while UniXcoder \cite{UniXcoder} and CodeT5+ \cite{CodeT5P} are fine-tuned for the same API recommendation task. 
The reliability of the API recommendation sequence is evaluated by the accuracy on accepted outputs (true accept rate), the recommendation rejection rate, and the false rejection rate, using the dataset released by Zhou et al. \cite{TheDevil}, constructed from the 50K-C dataset of compilable Java projects.
The evaluation results showed that NvRec achieved a remarkable accuracy of 83.8\% by dropping 80.7\% of API recommendation candidates, surpassing the best single model, MulaRec (46.3\%).

The contributions of this paper are as follows:
\begin{itemize}
    \item We propose \textbf{N-version API Recommendation (NvRec)}, a framework for generating reliable API sequence recommendations by filtering tail input cases and aggregating multiple inference results from diverse ML models.
    \item Through the empirical study, we demonstrate the effectiveness of TailAnalyzer, which improves the reliability of individual API recommendation models by filtering the tail cases. 
    \item We show that the three-version and five-version NvRec achieve significantly higher reliability than any single model on the extended 50K-C dataset. 
    In the best configuration, the three-version NvRec with CodeT5, MulaRec, and UniXcoder achieved \textbf{83.8\% true accept rate}, with a \textbf{80.7\% rejection rate} and a \textbf{32.2\% false rejection rate}.
    \item By comparing different filtering and output decision methods, we derive insights into selecting appropriate methods to achieve a higher true accept rate while reducing rejection rate and false rejection rate.
\end{itemize}

The remainder of this paper is organized as follows. Section II provides background on the API method sequence recommendation. Section III discusses related work. Section IV presents a performance comparison of ML Models for API method sequence recommendation. Section V details our proposed method. Section VI presents the evaluation results. Section VII discusses threats to validity. Finally, Section VIII concludes the paper and outlines future work.

\section{Background}
\subsection{API Usage and Challenges}
An API provides an interface for using the functionality of software components, such as operating systems, middleware, libraries, and frameworks. For example, the Java API \cite{JavaAPI}, the standard library for Java, provides functionality for file I/O, network communication, and GUI development. Developers can efficiently develop applications by using not only these standard libraries but also APIs provided by third parties \cite{HowToUseAPIDoc}.

Due to the large number of available libraries and frameworks, it is not easy to quickly identify the necessary functionalities and APIs during program development \cite{WhatAPIisHard, FieldStudyAPI}. A single library or framework often has multiple functionalities, making it time-consuming to find appropriate APIs. Furthermore, to use an API correctly, it is necessary to precisely understand the consequences of the API call. However, the documentation describing the overview and usage of APIs is not always sufficiently maintained \cite{LearningAPIisDifficult}. If the documentation is not well-maintained, developers may misuse the API and generate buggy code \cite{API-MisuseBug, API-Misuse2}. The use of APIs is essential for efficient software development, but quickly and accurately identifying the desired API remains a challenge \cite{DeepAPI}.

\subsection{API Method Sequence Recommendation}
\label{subsec:api_seq_rec_def}
In this study, we focus on API method sequence recommendation for programs written in Java. In object-oriented languages such as Java, APIs are concretely realized as methods defined in classes and interfaces. Therefore, we refer to each individual method provided by a library or framework as an \textit{API method}. An API method is represented as a string that connects the class name and the method name with a dot, such as \texttt{List.add} or \texttt{File.exists}. Furthermore, we define an \textit{API method sequence} as an ordered list of API methods invoked in accordance with dependencies and data flows to accomplish a specific functionality or task.

\begin{figure}[htbp]
 \centering
 \includegraphics[width=0.95\linewidth]{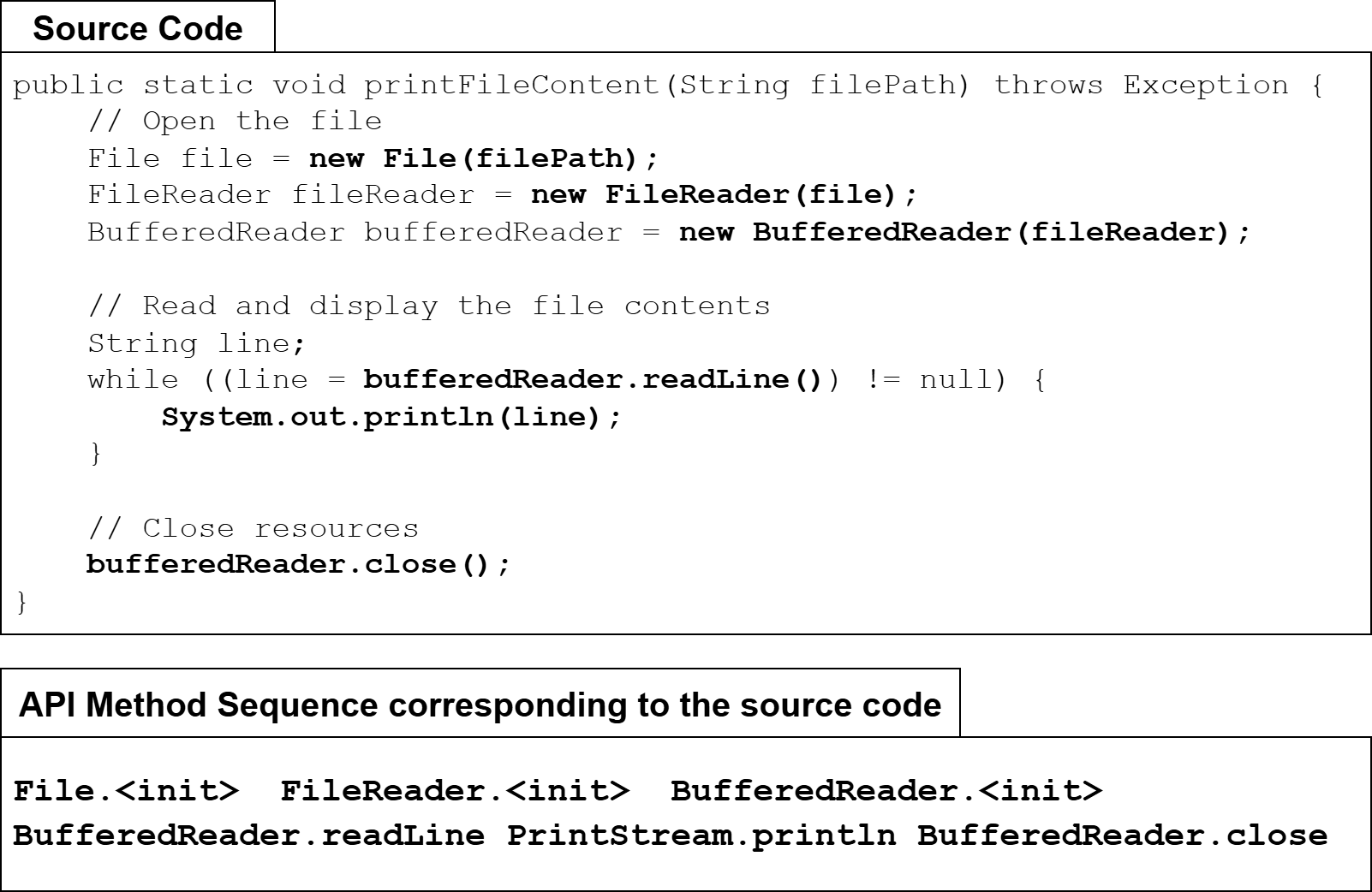}
 \caption{An example of source code and its corresponding API method sequence.}
 \label{fig:APISequenceSample}
\end{figure}

Fig.~\ref{fig:APISequenceSample} shows an example of source code and the corresponding API method sequence extracted from it. The source code shown at the top is a part of a Java program that reads the contents of a file and prints them to the console. The boldfaced parts in the code correspond to the major API method calls, including constructors. The bottom part shows the API method sequence obtained by listing the calls in execution order. For example, the instance creation \texttt{new File(filePath)} corresponds to \texttt{File.<init>}, and the method invocation \texttt{bufferedReader.readLine()} corresponds to \texttt{BufferedReader.readLine}. In this way, an API method sequence abstracts the implementation flow of source code as a sequence of API calls.

\begin{figure}[h] 
 \centering
 \includegraphics[width=0.75\linewidth]{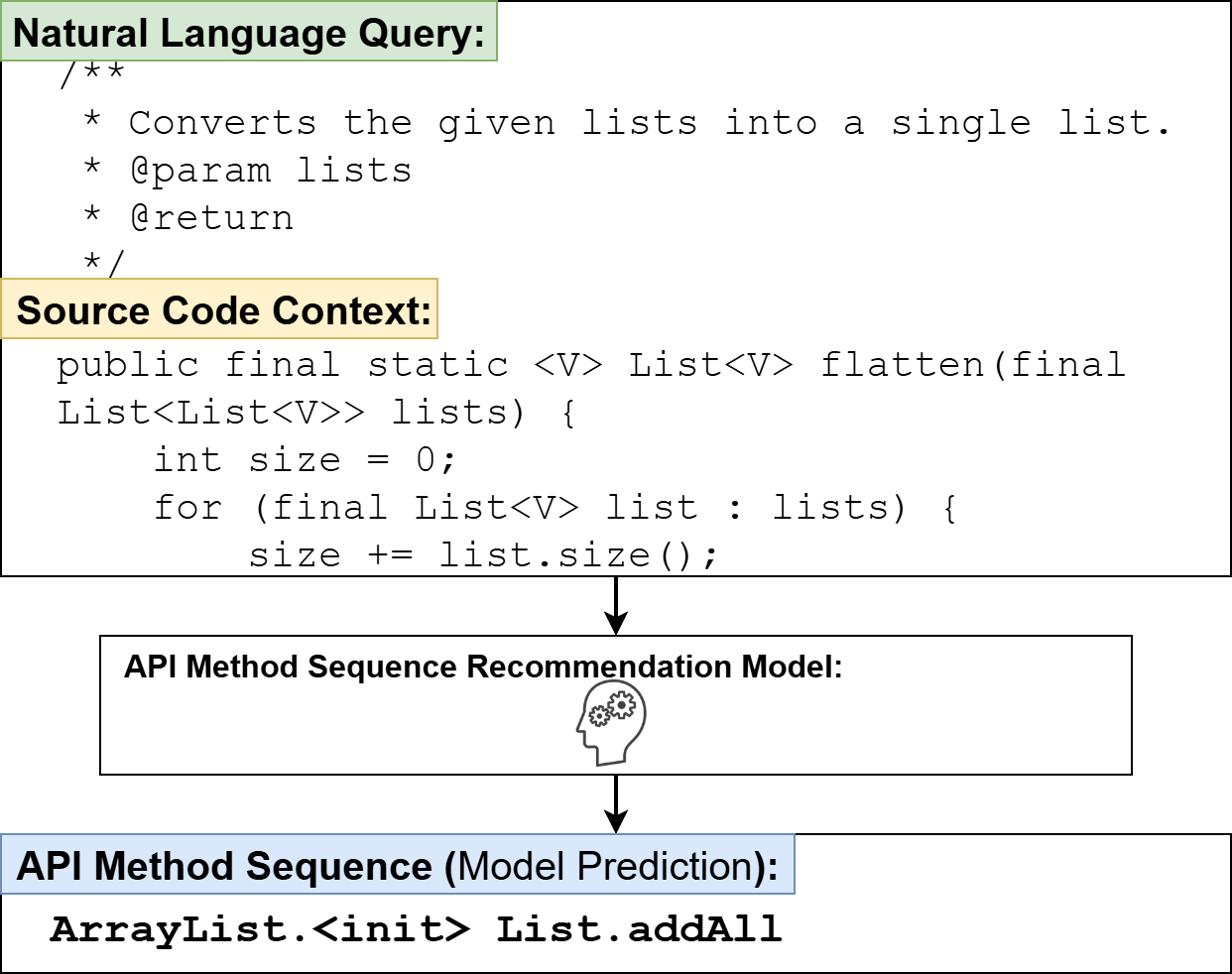}
 \caption{Overview of the API Method Sequence Recommendation Task}
 \label{fig:APISequenceRecommendTask}
\end{figure}

An API method sequence recommendation is a sequence-to-sequence task that takes a natural language query and a source code context as input and outputs a sequence of API methods to implement. Unlike single-API recommendation tasks, this task requires accurately inferring both the combination of API methods and their invocation order required to realize the target functionality. In this study, we follow the task definition and dataset introduced by Irsan et al.~\cite{MULAREC}. The dataset is constructed from a large collection of Java projects released by Martins et al.~\cite{50K-C}, where method information is extracted using the Eclipse JDT parser. Each data instance consists of a method declaration, the method body, and the associated Javadoc annotation.

The concrete definitions of inputs and outputs for the method sequence recommendation are as follows. The input comprises a natural-language query describing the intended functionality, represented by a Javadoc annotation, and a source code context under development, which includes the method declaration and the first three lines of the method body. Given these inputs, the model infers and outputs an API method sequence that should appear as the continuation of the source code. Fig.~\ref{fig:APISequenceRecommendTask} shows an overview of the API method sequence recommendation task in this study, where the green box represents the natural-language query, the yellow box represents the input source code context, and the blue box represents the ground-truth API method sequence to be predicted.

\subsection{Long-Tail Property and Head \& Tail API Methods}
\label{subsec:long_tail_api}

\begin{figure}[t] 
 \centering
 \includegraphics[width=\linewidth]{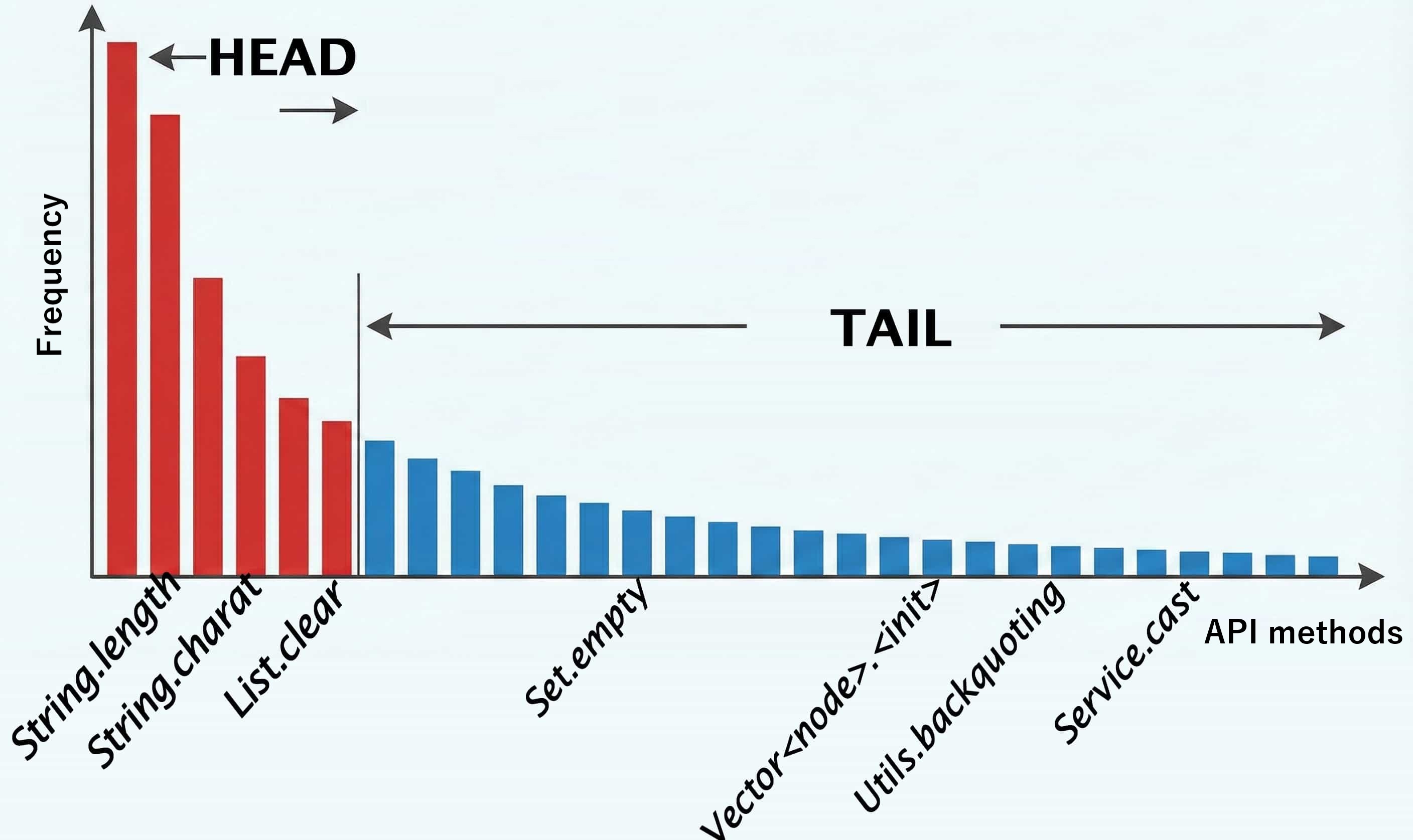}
 \caption{Long-tail distribution of API method frequency}
 \label{fig:Long-tailDistribution}
\end{figure}

In API method sequence recommendation, the distribution of API methods observed in real-world software projects exhibits a highly skewed pattern. While a small number of API methods are frequently used across many projects, the majority are used only in limited situations. The phenomenon is commonly known as the \textit{long-tail property}. As shown in Fig.~\ref{fig:Long-tailDistribution}, such a distribution is characterized by a \textit{head} region consisting of high-frequency items (e.g., top 50\% in the API usage counts) and a long \textit{tail} region composed of a large number of low-frequency items (i.e., remaining 50\% of the API usages).
In the context of API recommendation, general-purpose API methods such as \texttt{List.add} and \texttt{String.equals} are frequently used in a wide range of programs and therefore occupy the head of the distribution. In contrast, API methods that are specific to particular libraries, frameworks, or specialized functionalities are used in far fewer projects and consequently fall into the tail of the distribution.

The long-tail property poses a fundamental challenge for ML-based API method sequence recommendations~\cite{TheDevil, DDASR}. ML models are typically trained to minimize the average prediction error over the entire dataset, which biases them toward accurately recommending head API methods that are well represented in the training data. As a result, recommendations involving tail API methods, for which only limited training data are available, tend to suffer from lower accuracy. 

\section{Comparative Analysis of ML-based API Recommendation}\label{LLMによる評価}

\begin{table*}[t]
\centering
\caption{Comparison of the number and proportion of correct predictions for head API methods based on the threshold}
\label{事前調査ヘッドAPIメソッドの正解数表}
\begingroup
\renewcommand{\arraystretch}{1.2} 
\resizebox{\textwidth}{!}{
\begin{tabular}{c|ccccccc|c}
\hline
\diagbox{threshold $p$}{Model} & \makecell{CodeBERT\\only} & \makecell{CodeT5\\only} & \makecell{MulaRec\\only} & \makecell{CodeBERT\\\&CodeT5} & \makecell{CodeBERT\\\&MulaRec} & \makecell{CodeT5\\\&MulaRec} & \makecell{CodeBERT\\\&CodeT5\\\&MulaRec} & \makecell{Total} \\
\hline
10 & 46 (3.13\%) & 44 (2.99\%) & 84 (5.71\%) & 46 (3.13\%) & 108 (7.34\%) & 33 (2.24\%) & 1110 (75.46\%) & 1471 \\
20 & 102 (3.46\%) & 122 (4.14\%) & 189 (6.41\%) & 99 (3.36\%) & 210 (7.12\%) & 142 (4.82\%) & 2084 (70.69\%) & 2948 \\
30 & 152 (3.50\%) & 185 (4.26\%) & 298 (6.86\%) & 149 (3.43\%) & 316 (7.28\%) & 242 (5.57\%) & 3001 (69.10\%) & 4343 \\
40 & 195 (3.45\%) & 234 (4.15\%) & 404 (7.16\%) & 187 (3.31\%) & 420 (7.44\%) & 320 (5.67\%) & 3885 (68.82\%) & 5645 \\
50 & 236 (3.38\%) & 300 (4.29\%) & 496 (7.10\%) & 223 (3.19\%) & 500 (7.16\%) & 408 (5.84\%) & 4823 (69.04\%) & 6986 \\
60 & 282 (3.38\%) & 355 (4.26\%) & 598 (7.17\%) & 269 (3.23\%) & 591 (7.09\%) & 518 (6.21\%) & 5725 (68.66\%) & 8338 \\
70 & 320 (3.31\%) & 416 (4.30\%) & 689 (7.12\%) & 322 (3.33\%) & 668 (6.90\%) & 590 (6.10\%) & 6672 (68.95\%) & 9677 \\
80 & 347 (3.16\%) & 484 (4.40\%) & 807 (7.34\%) & 362 (3.29\%) & 746 (6.79\%) & 671 (6.11\%) & 7572 (68.91\%) & 10989 \\
90 & 375 (3.08\%) & 552 (4.54\%) & 916 (7.53\%) & 410 (3.37\%) & 800 (6.57\%) & 749 (6.15\%) & 8368 (68.76\%) & 12170 \\
\hline
\end{tabular}
}
\endgroup
\end{table*}

\begin{table*}[t]
\centering
\caption{Number and proportion of correct predictions for tail API methods based on the threshold}
\label{事前調査テールAPIメソッドの正解数表}
\begingroup
\resizebox{\textwidth}{!}{
\begin{tabular}{c|ccccccc|c}
\hline
\diagbox{threshold $p$}{Model} & \makecell{CodeBERT\\only} & \makecell{CodeT5\\only} & \makecell{MulaRec\\only} & \makecell{CodeBERT\\\&CodeT5} & \makecell{CodeBERT\\\&MulaRec} & \makecell{CodeT5\\\&MulaRec} & \makecell{CodeBERT\\\&CodeT5\\\&MulaRec} & \makecell{Total} \\
\hline
10 & 357 (3.00\%) & 583 (4.89\%) & 943 (7.92\%) & 416 (3.49\%) & 748 (6.28\%) & 799 (6.71\%) & 8065 (67.71\%) & 11911 \\
20 & 301 (2.88\%) & 505 (4.84\%) & 838 (8.03\%) & 363 (3.48\%) & 646 (6.19\%) & 690 (6.61\%) & 7091 (67.96\%) & 10434 \\
30 & 251 (2.78\%) & 442 (4.89\%) & 729 (8.07\%) & 313 (3.46\%) & 540 (5.97\%) & 590 (6.53\%) & 6174 (68.30\%) & 9039 \\
40 & 208 (2.69\%) & 393 (5.08\%) & 623 (8.05\%) & 275 (3.55\%) & 436 (5.64\%) & 512 (6.62\%) & 5290 (68.37\%) & 7737 \\
50 & 167 (2.61\%) & 327 (5.11\%) & 531 (8.30\%) & 239 (3.74\%) & 356 (5.57\%) & 424 (6.63\%) & 4352 (68.04\%) & 6396 \\
60 & 121 (2.40\%) & 272 (5.39\%) & 429 (8.51\%) & 193 (3.83\%) & 265 (5.25\%) & 314 (6.23\%) & 3450 (68.40\%) & 5044 \\
70 & 83 (2.24\%) & 211 (5.70\%) & 338 (9.12\%) & 140 (3.78\%) & 188 (5.07\%) & 242 (6.53\%) & 2503 (67.56\%) & 3705 \\
80 & 56 (2.34\%) & 143 (5.98\%) & 220 (9.19\%) & 100 (4.18\%) & 110 (4.60\%) & 161 (6.73\%) & 1603 (66.99\%) & 2393 \\
90 & 28 (2.31\%) & 75 (6.19\%) & 111 (9.16\%) & 52 (4.29\%) & 56 (4.62\%) & 83 (6.85\%) & 807 (66.58\%) & 1212 \\
\hline
\end{tabular}
}
\endgroup
\end{table*}

Existing studies addressed the challenge of long-tail distribution in ML-based API sequence recommendations~\cite{TheDevil, DDASR}.
However, these studies did not discuss how different ML models produce different recommendation results for tail API methods. Even if one model recommends an incorrect API method, another model may recommend the correct one. The diversity of ML models' recommendation capabilities can be exploited to improve the reliability of API sequence recommendations.

\begin{figure}[t]
\centering
\includegraphics[width=0.5\linewidth]{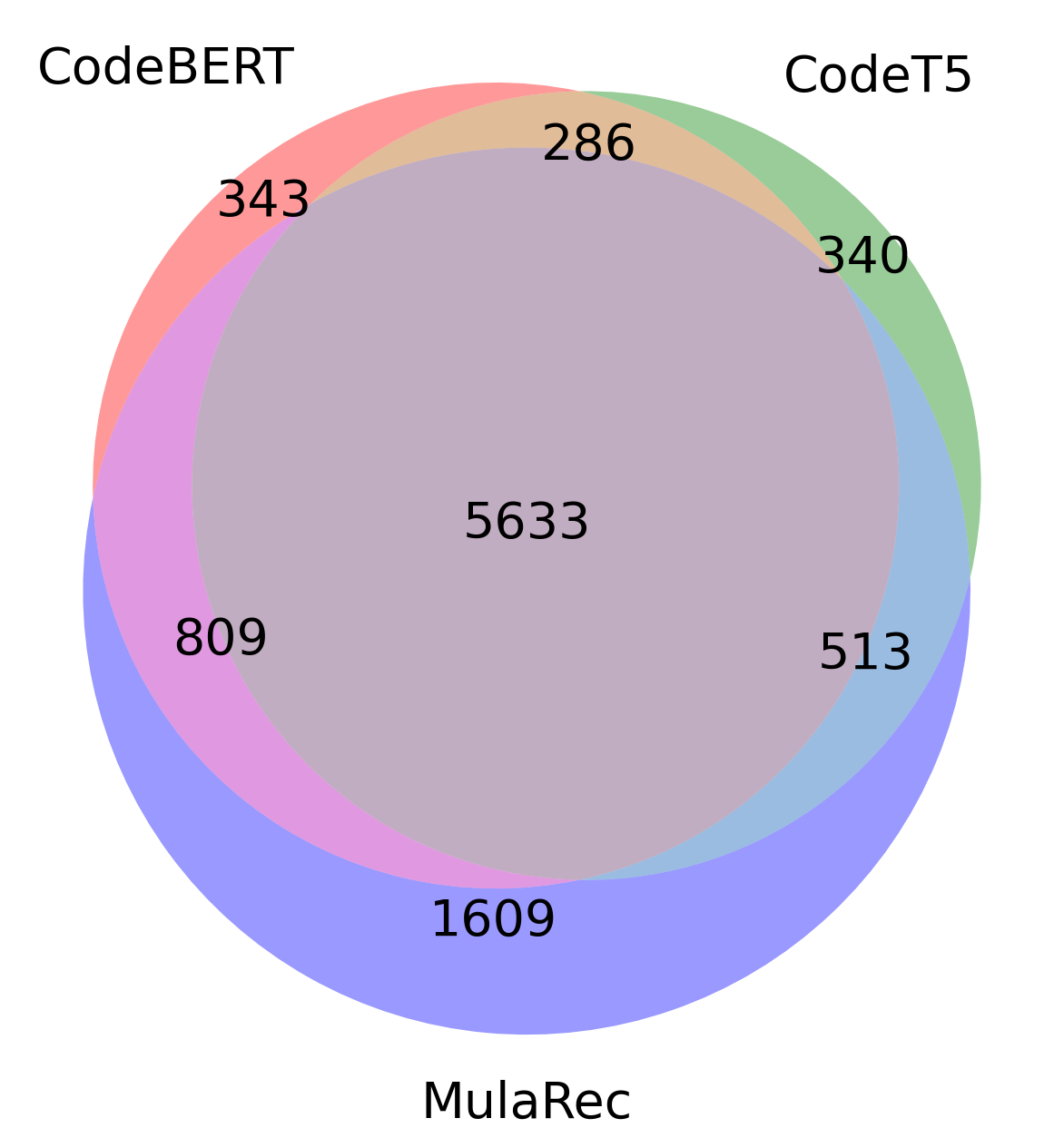}
\caption{The number of correct predictions by three ML models.}
\label{各LLMの正解状況}
\end{figure}

Motivated by this, we conducted a comparative evaluation of API sequence recommendation models, namely CodeBERT \cite{CodeBERT}, CodeT5 \cite{CodeT5}, and MulaRec \cite{MULAREC}. These ML models are selected because their long-tail properties were examined by Zhou et al. \cite{TheDevil}, and their evaluation datasets are available for reproducing the results. This dataset contains 18,500 entries, each consisting of a natural-language query, a source code context, and a ground-truth API method sequence. The accuracy of the models on this input data for API method sequences is 38.2\% for CodeBERT, 36.6\% for CodeT5, and 46.3\% for MulaRec. A breakdown of the correctly predicted API method sequences is shown in Fig. \ref{各LLMの正解状況}.
Among the three models, MulaRec produced the most correct predictions.
Moreover, 343 samples are correctly predicted only by CodeBERT. Meanwhile, we can also see that 340 samples were correctly predicted only by CodeT5, which is the weakest model. The results indicate that model diversity exists in the API method recommendation capabilities.

To further investigate the differences in the inference capabilities of each ML model for head and tail API methods, we analyzed them by varying the threshold that separates the head and tail. Table \ref{事前調査ヘッドAPIメソッドの正解数表} and Table \ref{事前調査テールAPIメソッドの正解数表} show the number of correct predictions and their proportions for head and tail API methods, respectively. The threshold in these tables is the criterion for classifying the entire dataset into "top $p$\% head API methods" and "bottom $(100-p)$\% tail API methods." For example, if $p=10$, the top 10\% of the data is classified as head API methods, and the remaining 90\% is classified as tail API methods. Based on the classification, we measured the number of correct predictions for head and tail API methods separately. The number in each cell shows the number of correct predictions by the adopted models and the proportion of all the correctly predicted samples. For example, when the tail threshold $p=90$, the unique correct predictions made by CodeBERT for the head API methods are counted as 375, which is 3.08\% of 12170, representing the total number of samples correctly predicted by any one of the models.

By comparing Tables \ref{事前調査ヘッドAPIメソッドの正解数表} and \ref{事前調査テールAPIメソッドの正解数表}, we can discern the differences in the recommendation capabilities of the ML models for head and tail API methods. First, looking at the proportion of correct predictions made only by CodeBERT, it is consistently over 3.05\% for head APIs, but it is 3\% or less for tail APIs. Therefore, it is considered that CodeBERT solely has a weaker recommendation capability for tail APIs than for head APIs. On the other hand, the proportion of correct predictions made only by CodeT5 or only by MulaRec is higher for tail APIs than for head APIs. 
Note that this result does not imply that recommendations of tail APIs are more reliable than those of head APIs, rather simply tells that the unique prediction capabilities of CodeT5 and MulaRec are more apparent in tail APIs. The unique prediction capabilities are model-dependent and do not consistently appear in the head or tail of the distributions.
Moreover, across all three models, the proportion of correct predictions for head APIs is very high at a lower threshold (e.g., 75.46\% when $p=10$). This indicates that correct recommendations of API methods at the head of the distribution are relatively easy for any model.

In summary, the comparative analysis gives the following insights: 1) the API recommendation capability of ML models differs for head and tail APIs, 2) there is room to leverage the unique recommendation capabilities of different ML models 
and 3) it is likely that all ML models correctly predict in common for head APIs. Based on this observation, we propose an API recommendation method that leverages the diverse capabilities of ML models in their recommendations.

\begin{figure*}[htbp]
\centering
\includegraphics[width=\textwidth]{images/Overview_of_MyApproachNew_ENG.jpg}
\caption{Overview of NvRec. The main process of NvRec is divided into two steps: 1) model profile step that generates the model profile based on the API recommendation performance of the ML models, and 2) inference step that recommends API method sequence for given inputs based on the candidates generated from multiple ML models, or rejects the outputs.}
\label{提案手法の全体像}
\end{figure*}

\begin{table*}[htbp]
\centering
\caption{An example of a model profile that records the history of api method predictions by different ML models.}
\label{モデルプロファイル}
\resizebox{\textwidth}{!}{
\begin{tabular}{lllrrrrrrrr}
\toprule
\multicolumn{1}{c}{\multirow{2}{*}{\textbf{API method name}}} & \multicolumn{1}{c}{\multirow{2}{*}{\textbf{Frequency}}} & \multicolumn{1}{c}{\multirow{2}{*}{\textbf{Tail}}} & \multicolumn{2}{c}{\textbf{Model 1 Performance}} & \multicolumn{2}{c}{\textbf{Model 2 Performance}} & \multicolumn{1}{c}{\dots} & \multicolumn{2}{c}{\textbf{Model N Performance}} \\ 
\cmidrule(lr){4-5} \cmidrule(lr){6-7} \cmidrule(lr){9-10}
\multicolumn{1}{c}{} & \multicolumn{1}{c}{} & \multicolumn{1}{c}{} & \multicolumn{1}{r}{\textbf{Recommendations}} & \multicolumn{1}{r}{\textbf{Correct}} & \multicolumn{1}{r}{\textbf{Recommendations}} & \multicolumn{1}{r}{\textbf{Correct}} & \multicolumn{1}{c}{\dots} & \multicolumn{1}{r}{\textbf{Recommendations}} & \multicolumn{1}{r}{\textbf{Correct}} \\ 
\midrule
String.length & 227 & H & 245 & 144 & 228 & 130 & \dots & 251 & 145 \\
String.cast & 223 & H & 205 & 181 & 172 & 171 & \dots & 210 & 190 \\ 
String.equals & 208 & H & 254 & 110 & 167 & 89 & \dots & 236 & 123 \\ 
ArrayList.add & 100 & H & 121 & 66 & 91 & 56 & \dots & 89 & 62 \\
node.cast & 16 & H & 13 & 13 & 17 & 14 & \dots & 18 & 15 \\ 
: & : & : & : & : & : & : & \dots & : & : \\
Random.nextFloat & 5 & T & 3 & 3 & 3 & 3 & \dots & 5 & 3 \\
Tile.getPolygon & 0 & T & 0 & 0 & 1 & 0 & \dots & 0 & 0 \\
Vector.notifyall & 0 & T & 1 & 0 & 0 & 0 & \dots & 0 & 0 \\
\bottomrule
\end{tabular}
}
\end{table*}

\section{Proposed method}
\subsection{Overview of the proposed method}
We propose N-version API recommendation (NvRec) that aims to improve the reliability of API method sequence recommendations by filtering unreliable recommendations using multiple versions of ML models. Instead of designing new ML models and training them on large code datasets, NvRec provides a framework for leveraging pre-trained ML models to generate more reliable API recommendations.
To suppress unreliable outputs, NvRec predicts tail cases for given inputs and discards them, while multi-version inference is applied for the head case to generate reliable recommendation results.

The overview of NvRec is shown in Fig. \ref{提案手法の全体像}.
The main process of NvRec is divided into two steps: a \textit{model profiling} step and an \textit{inference} step. In the \textit{model profiling} step, a model profile is generated from the records of the API recommendations produced by different ML models on a profiling subset of the provided test dataset.
Each record of the model profile corresponds to a specific API and includes the tail flag and recommendation histories for each ML model. The tail flag is set when the API method is at the tail of the distribution in terms of the appearance frequency. To judge the tail, the API methods included in the training dataset are sorted in descending order of the number of occurrences. Then, the API methods appeared in the top $p$\% of the cumulative frequency are recorded as head, and the others are recorded as tail.

In the \textit{inference} step, we introduce \textit{Tail Analyzer} that judges if the given natural language query and source code context fall into the tail sample. If \textit{Tail Analyzer} classifies the input as a tail sample, we discard the request because the resulting API recommendations are expected to be unreliable. NvRec performs N-version API recommendation only when \textit{Tail Analyzer} classifies the input as a head sample (i.e., judged as top $p$\% samples). The candidates generated by N-version ML models are filtered based on the model profile to eliminate unreliable API method sequence candidates. 
The final output is then determined by applying a majority voting scheme to the remaining candidates. 
If the voting process fails to reach a consensus, NvRec either selects a promising candidate by consulting the model profile or rejects the recommendation to avoid recommending unreliable outputs.

By filtering tail cases and exploiting the diversity of N-version ML models with the help of the model profile, NvRec can reject unreliable API recommendations, yielding more reliable API method sequence recommendations.

\subsection{Model Profile}

The model profile is a critical component to leverage the diverse capabilities of N-version ML models. An example of a model profile is shown in Table \ref{モデルプロファイル}. Using the profiling dataset, a model profile is prepared for each ML model by tracking the API methods that appear in its recommendations. For each API method, the number of times it appeared in the recommendations and the number of correct recommendations are recorded. From the counts of individual APIs in the profiling dataset, we can derive the empirical distributions of API methods within the correct API method sequences. Let $n$ be the total number of API methods in the dataset, and $n_i$ be the number of appearances for the specific API method $i$. In the profile, we record $n_i$ as the frequency of each API method. Assuming that API methods are ordered in a descending order of $n_i$, the tail flag for the API method $i$ is set by
\begin{equation}
    \textit{tail\_flag(i)} = \begin{cases}
                            H & \frac{1}{n}\sum_{k=1}^{i} n_k \leq \frac{p}{100}, \\
                            T & otherwise.
                            \end{cases}
\end{equation} 
where $p$ is the tail threshold. The API methods with a lower frequency are categorized as tail API methods ($T$). 

Next, the performance of individual ML models is evaluated with the profile dataset. By collecting the recommended API methods, the \textit{number of recommendations} is recorded (See Table \ref{モデルプロファイル}). Then, comparing the results with the correct API method sequence, the profile also records the \textit{number of correct predictions}. While the correctness is checked by API method sequences, the profile simply counts the number of API methods that appear in the correct API recommendations. The number of recommendations and the number of correct predictions are counted for individual models.

As shown in the example in Table \ref{モデルプロファイル}, a frequently occurring head API method, such as \textit{String.length}, yields a high number of recommendations and correct predictions across all models. On the other hand, a rarely occurring tail API method such as \textit{Random.nextfloat} shows a relatively high accuracy, although the number of recommendations is small.

The statistics recorded in the profile represent the performance characteristics of individual ML models for each API method. Therefore, the model profile provides a cheat sheet for selecting reliable models for a given API method in the inference step. 

\subsection{Tail Analyzer}
In the \textit{inference} step, we do not know if the input natural language query and source code context are in a tail case. Therefore, we introduce  \textit{Tail Analyzer} to predict whether the input belongs to the tail case. Tail Analyzer is trained on the same dataset as the API method sequence recommendation task; however, the learning objective and labels differ. Following Zhou et al.~\cite{TheDevil}, we use the binary head/tail labels provided in their dataset, which are defined based on the frequency of API methods in the training data. Specifically, an input instance is labeled a tail case if its ground-truth API method sequence contains API methods categorized as tail APIs under the frequency-based criterion. For this task, the input is a single string that concatenates the natural language query and the source code context, and the output is a binary flag indicating whether the input corresponds to a tail case. If the input is predicted as the head case, the recommended API method sequence is likely to consist solely of head API methods, yielding reliable API recommendations. Thus, in this case, we use N-version ML inferences to make a careful API sequence recommendation. Conversely, if the input is predicted as a tail case, the recommended API method sequence likely consists of tail APIs, leading to unreliable API recommendations. As NvRec aims to reduce unreliable API recommendations, we drop the request and do not proceed with the next step in the tail case.

\subsection{Multi-Version Inference}
In the multi-version inference, $N (\geqq 2)$ different versions of ML models independently generate API method sequence candidates for a given natural language query and source code context. Like N-version programming for fault-tolerant software \cite{N-VERSION}, the independent execution of multiple components for the same input is essential to ensure diversity. As a result of the multi-version inference, individual ML models may recommend different API method sequences. We need a decision rule to make the final API recommendation output from the given candidates.

\subsection{Filtering}
After obtaining multi-version inference results, we introduce a filtering step to ensure the quality of candidates based on the model profile before making the final decision.
Specifically, one of the following filters is applied.
\vspace{1ex}

\noindent\textbf{Recorded filtering (R-Filter):} The filter checks whether all API methods included in the recommended sequence are recorded in the model profile. If any of the API methods do not exist in the model profile, the recommended sequence is deemed unreliable and dropped from the candidates.
\vspace{1ex}

\noindent\textbf{Head filtering (H-Filter):} The filter checks the tail flag of all the API methods included in the candidate sequence. If one or more tail API methods are included, the recommended sequence is deemed unreliable and is dropped from the candidates.
\vspace{1ex}

\noindent\textbf{Recorded and head filtering (RH-Filter):} The filter checks both the past prediction record and the tail flag for the API methods in the recommended sequence. This filter is the combination of the R-filter and the H-filter. If any one of the API methods does not exist in the record or is flagged as a tail, the recommended sequence is dropped from the candidates.

\subsection{Output Decision}
After filtering unreliable recommendation candidates, the final API method sequence recommendation result is selected by a decision rule. If more than one API method sequence candidate remains, the majority vote of the candidates is applied. If the majority reaches consensus on the same API method sequence, the result is selected as the final output. On the other hand, if the majority vote does not reach consensus on a specific API method sequence, the final output is determined by applying one of the following decision methods.
\vspace{1ex}

\noindent\textbf{Simple rejection:} If there is no API method sequence candidate that reaches a majority, all the candidates are dropped, leading to no output from the system.
\vspace{1ex}

\noindent\textbf{Reliability-score based decision (score-based):} For each API sequence recommendation candidate, the reliability score is computed based on the model profile. The reliability score is defined as the geometric mean of the accuracy rates of all the API methods in the sequence.
\begin{equation}
    Score(S_i) = \left( \prod_{m \in S_i} P(m|M_i) \right)^{1/|S_i|}
\end{equation}
where $P(m|M_i)$ represents the accuracy rate on the model profile of model $M_i$ for API method $m$, and $|S_i|$ represents the number of APIs in the sequence. The geometric mean ensures that the score resets to zero if any model has never made a successful recommendation. If the best API method sequence, in terms of reliability score, exceeds the threshold $\theta$, it is taken as the final output. If the best score does not reach $\theta$, all the recommendations are discarded.
\vspace{1ex}

\noindent\textbf{Using the best model's output (best model):} The API sequence recommended by the best performance model that achieves the highest accuracy rate is adopted and used as the final output. If the candidates do not include the inference result by the best model, all other candidates are dropped. Similar to the score-based approach, the reliability score of the best model's recommendation is also computed, and the sequence is adopted as the final output only if its score is greater than or equal to the threshold $\theta$.

\section{Evaluation}
\subsection{Research Questions}
To evaluate the effectiveness of NvRec, we set the following research questions.\\
\noindent\textbf{RQ1.} How effective is \textit{Tail Analyzer} alone in improving the reliability of API method sequence recommendation?\\
\noindent\textbf{RQ2.} How does NvRec improve the reliability of API method sequence recommendation?\\
\noindent\textbf{RQ3.} How is the reliability of API method sequence recommendation affected by different output decision rules?\\
\noindent\textbf{RQ4.} Which filtering method is the most effective for improving the reliability of API method sequence recommendation?

For reliability evaluation, we primarily use the rate of exact matches with the ground truth among the accepted outputs, defined as the true accept rate (TAR).
Since NvRec proactively reduces unreliable recommendations through filtering, we also evaluate the rejection rate (RR), which represents the proportion of rejected requests over all inputs.
Furthermore, to assess how many correct recommendations are mistakenly discarded by filtering, we introduce the false rejection rate (FRR).
The formal definitions of these metrics are provided in Section~\ref{subsec:evaluation_method}.

\subsection{Experimental Setup}\label{subsec:experimental_setup}
For the implementation of NvRec, we adopted publicly available pretrained ML models for the API method sequence recommendation task. Specifically, we used five models: CodeBERT, CodeT5, MulaRec, UniXcoder, and CodeT5+. These five models constitute the components of the five-version NvRec and are also used in the three-version NvRec configurations evaluated in the subsequent evaluation section. For CodeBERT, CodeT5, and MulaRec, we directly used the publicly released fine-tuned models without any additional training. In contrast, UniXcoder and CodeT5+ were further fine-tuned using the same dataset employed in the comparative analysis of the other models. The fine-tuning procedure followed the same settings as those used for CodeBERT, CodeT5, and MulaRec, except that the number of training epochs was set to 15.
All the models used in the evaluation are included in our reproducible research package.
\footnote{https://doi.org/10.5281/zenodo.17346142}

For the input and output length constraints, we followed the default configurations of each model implementation. CodeBERT and CodeT5 were originally trained with a maximum input length of 256 tokens and a maximum output length of 100 tokens. MulaRec allows longer outputs, supporting up to 256 output tokens. UniXcoder and CodeT5+ also accept a maximum input length of 256 tokens and a maximum output length of 100 tokens.

In addition to the recommendation models, we implemented \textit{Tail Analyzer} to detect tail cases at the beginning of the inference step. \textit{Tail Analyzer} is a binary classification model, and we trained separate instances of Tail Analyzer based on existing API method sequence recommendation models.
The training dataset for \textit{Tail Analyzer} consists of 295,995 samples, where each sample is a pair of a single string, created by concatenating a natural language query and its source code context, and a binary label (head: 0, tail: 1). Following Zhou et al.~\cite{TheDevil}, we adopt the binary head/tail labels provided in their dataset, in which the boundary between head and tail cases is defined based on API method frequency, with the top 50\% of the samples categorized as head cases. The definition is directly inherited from the dataset and is not newly introduced in this study. 
Although we conducted a sensitivity analysis to investigate the impact of the tail-head split ratio (i.e., $p\%$), there was no notable difference in the main conclusions. Thus, following the previous study, we set $p=50$ as the tail-split ratio for the following evaluations.

\subsection{Evaluation Methodology}
\label{subsec:evaluation_method}
To evaluate the effectiveness of \textit{Tail Analyzer}, we first compare the performance of individual ML models with and without applying \textit{Tail Analyzer}.
We then evaluate three-version and five-version configurations of NvRec under different combinations of filtering methods (R-Filter, H-Filter, and RH-Filter) and output decision rules (Simple Rejection, Score-based, and Best model).

The adopted evaluation metrics are defined as follows:
\begin{align}
    \text{\textit{Rejection Rate (RR)}} 
    &= 1 - \frac{\text{\textit{\# of Total outputs}}}{\text{\textit{\# of Total inputs}}}, \\
    \notag \\
    \text{\textit{True Accept Rate (TAR)}} 
    &= \frac{\text{\textit{\# of Correct outputs}}}{\text{\textit{\# of Total outputs}}}, \\
    \notag \\
    \text{\textit{False Rejection Rate (FRR)}} 
    &= \frac{\text{\textit{\# of Rejected Correct outputs}}}{\text{\textit{\# of Rejected outputs}}}.
\end{align}
\vspace{1ex}

Note that in this paper, the term \textit{accuracy} is used only for evaluating \textit{Tail Analyzer} as a binary classifier, whereas the recommendation performance is evaluated primarily by TAR, RR, and FRR. RR represents the proportion of input instances that are rejected by \textit{Tail Analyzer}, the filtering step, or the output decision rule.
TAR measures the proportion of accepted API method sequence recommendations that exactly match the ground-truth sequence, reflecting the reliability of the system among the outputs that are actually returned to users.
For a single ML model without Tail Analyzer, 
TAR is equal to the accuracy since it does not have rejections.
The FRR denotes the proportion of rejected outputs that could have resulted in a correct recommendation if they had not been discarded.
A \textit{rejected correct output} refers to a recommendation that is filtered out despite being potentially correct according to the majority vote of candidate sequences.
Since such outputs are unnecessarily removed, a lower FRR is preferable.

In the Score-based and Best model output decision rules, the rejection threshold $\theta$ is set to 0.9. 
This is because the objective of our approach is to recommend only highly reliable API method sequences, and a score closer to 1 indicates a higher likelihood that the candidate sequence is correct.
Since MulaRec achieves the highest TAR among the individual ML models considered in this study, we use MulaRec as the best model in the Best model decision rule.

\subsection{Results}
\label{subsec:evaluation_result}

\subsubsection{Effectiveness of Tail Analyzer}
\label{subsubsec:tail_analyzer_effectiveness}

We first evaluate the effectiveness of  \textit{Tail Analyzer} constructed from CodeBERT, CodeT5, MulaRec, UniXcoder, and CodeT5+.
Table~\ref{tab:tail_analyzer_performance} reports the classification accuracy of each \textit{Tail Analyzer} on the evaluation dataset.
All models achieve relatively high accuracy, indicating that the binary head/tail classification task can be learned effectively.
Among them, UniXcoder achieves the highest accuracy of 85.82\%.
Therefore, in the subsequent experiments, we adopt UniXcoder as the backbone model for  \textit{Tail Analyzer}.

\begin{table}[h]
\centering
\caption{Classification accuracy of the \textit{Tail Analyzer}}
\label{tab:tail_analyzer_performance}
\begin{tabular}{lr}
\hline
\textbf{Model} & \textbf{Accuracy (\%)} \\ \hline
CodeBERT & 84.29 \\
CodeT5 & 85.28 \\
MulaRec & 83.55 \\ 
UniXcoder & $\mathbf{85.82}$ \\ 
CodeT5+ & 84.25 \\ \hline
\end{tabular}
\end{table}

Next, to answer RQ1, we investigate the impact of applying the \textit{Tail Analyzer} on the reliability of API method sequence recommendation.
Table~\ref{tab:tail_analyzer_impact} compares the TAR of individual recommendation models with and without \textit{Tail Analyzer}.

\begin{table}[h]
\centering
\caption{Impact of Tail Analyzer on the TAR of individual models}
\label{tab:tail_analyzer_impact}
\begingroup
\renewcommand{\arraystretch}{1.2}
\begin{tabular}{lrr}
\hline
\textbf{Model} & \textbf{w/o Tail Analyzer} & \textbf{w/ Tail Analyzer} \\ \hline
CodeBERT  & 38.1 & 45.7 \\
CodeT5    & 36.8 & 42.3 \\
MulaRec   & $\mathbf{46.3}$ & $\mathbf{52.2}$ \\
UniXcoder & 40.0 & 45.7 \\
CodeT5+   & 39.1 & 44.8 \\ \hline
\end{tabular}
\endgroup
\end{table}

In the baseline setting without \textit{Tail Analyzer}, all input instances are accepted for recommendation, resulting in a RR of 0\%.
When \textit{Tail Analyzer} is applied, 51.6\% of the inputs are classified as tail cases and rejected in advance.
As shown in Table~\ref{tab:tail_analyzer_impact}, the prior rejection consistently improves the TAR for all individual models.
For example, the TAR of MulaRec increases from 46.3\% to 52.2\%, and that of CodeBERT improves from 38.1\% to 45.7\%.
These results indicate that \textit{Tail Analyzer} functions effectively across different model architectures by rejecting tail cases that are more likely to lead to incorrect recommendations, thereby improving the reliability of accepted API method sequence recommendations.

\vspace{2ex}
\noindent\begin{tikzpicture}
\node [draw,
rounded corners=5pt,
fill=gray!10,
inner sep=8pt,
align=justify,
text width=0.95\columnwidth] {
\textbf{Answer to RQ1:} \textit{Tail Analyzer} improves the reliability of API method sequence recommendation by proactively rejecting input instances that are likely to result in incorrect recommendations, thereby increasing the TAR of the accepted outputs.};
\end{tikzpicture}

\subsubsection{Reliability Improvement by NvRec}
\label{subsubsec:reliability_improvement}
We evaluate how NvRec improves the reliability of API method sequence recommendation by combining multiple ML models, filtering methods, and output decision rules.
We first analyze three-version NvRec configurations and then compare three-version and five-version NvRec.

\begin{table}[t]
\centering
\caption{Ten three-model combinations ranked by TAR under H-Filter with Simple Rejection}
\label{tab:top10_combinations}
\resizebox{\columnwidth}{!}{%
\begin{tabular}{p{4.8cm} r r r}
\hline
\textbf{Models} & \textbf{TAR} & \textbf{RR} & \textbf{FRR} \\
\hline
CodeT5, MulaRec, UniXcoder   & $\mathbf{83.8}$ & 80.7 & 32.2 \\
CodeT5, MulaRec, CodeT5+    & 82.8            & 80.6 & 31.7 \\
CodeBERT, MulaRec, UniXcoder& 77.3            & 78.5 & 32.5 \\
CodeBERT, MulaRec, CodeT5+  & 76.8            & 78.7 & 31.9 \\
CodeBERT, CodeT5, MulaRec   & 78.4            & 80.2 & 29.6 \\
CodeBERT, CodeT5, UniXcoder & 78.2            & 80.3 & 28.9 \\
CodeBERT, CodeT5, CodeT5+   & 77.8            & 80.3 & 28.8 \\
MulaRec, UniXcoder, CodeT5+ & 72.7            & 78.7 & 30.5 \\
CodeT5, UniXcoder, CodeT5+  & 74.0            & 79.6 & 29.1 \\
CodeBERT, UniXcoder, CodeT5+& 69.4            & 77.8 & 29.6 \\
\hline
\end{tabular}%
}
\end{table}

Among the five ML models considered in this study, choosing three models yields ten possible combinations.
After evaluating all the possible combinations, we observed that
the configuration using H-Filter with Simple Rejection achieved the highest TAR regardless of combinations.
Thus, we report the ten three-model combinations ranked by the TAR under H-Filter + Simple Rejection in Table~\ref{tab:top10_combinations}.
Among the ten configurations, the highest TAR is achieved by the combination of CodeT5, MulaRec, and UniXcoder.
This configuration benefits from the stronger ML models, MulaRec and UniXcoder. Although CodeT5 underperforms CodeBERT in standalone TAR, CodeT5 often exhibits error patterns that differ from those of MulaRec and UniXcoder.
Consequently, majority voting is more likely to cancel out model-specific errors, improving the reliability of the accepted outputs.


Next, we compare the detailed performance of NvRec between a representative three-version configuration and the five-version configuration.
For the three-version NvRec, we report the results of the best-performing three-model combination (CodeT5, MulaRec, UniXcoder) under all combinations of filtering methods and output decision rules.
For the five-version NvRec, we report the results under the same set of filters and decision rules.

\begin{table}[t]
\caption{Evaluation Results of Three-version NvRec (CodeT5, MulaRec, UniXcoder)}
\label{tab:eval_ct5_mr_ux}
\centering
\begingroup
\resizebox{0.9\columnwidth}{!}{%
\begin{tabular}{llrrr}
\toprule
\textbf{Filter} &
\textbf{Decision Rule} &
\textbf{TAR (\%)} &
\textbf{RR (\%)} &
\textbf{FRR (\%)} \\
\midrule
R-Filter  & Simple Rejection & 82.9 & 74.8 & 28.5 \\
          & Score-based      & 77.1 & 72.4 & 29.2 \\
          & Best Model       & 80.3 & 73.8 & 28.8 \\
\midrule
H-Filter  & Simple Rejection & $\mathbf{83.8}$ & 80.7 & 32.2 \\
          & Score-based      & 81.5 & 80.0 & 32.5 \\
          & Best Model       & 82.7 & 80.4 & 32.3 \\
\midrule
RH-Filter & Simple Rejection & 83.7 & 83.2 & 33.8 \\
          & Score-based      & 81.0 & 82.5 & 34.1 \\
          & Best Model       & 82.4 & 82.9 & 33.9 \\
\bottomrule
\end{tabular}%
}
\endgroup
\end{table}

\begin{table}[t]
\caption{Evaluation Results of Five-version NvRec}
\label{tab:eval_results_all_five}
\centering
\begingroup
\resizebox{0.9\columnwidth}{!}{%
\begin{tabular}{llrrr}
\toprule
\textbf{Filter} &
\textbf{Decision Rule} &
\textbf{TAR (\%)} &
\textbf{RR (\%)} &
\textbf{FRR (\%)} \\
\midrule
R-Filter & Simple Rejection & $\mathbf{67.1}$ & 67.2 & 27.5 \\
         & Score-based      & 62.0 & 64.1 & 28.9 \\
         & Best Model       & 65.7 & 66.4 & 27.9 \\
\midrule
H-Filter & Simple Rejection & 60.8 & 72.4 & 33.1 \\
         & Score-based      & 59.4 & 71.6 & 33.5 \\
         & Best Model       & 60.3 & 72.2 & 33.2 \\
\midrule
RH-Filter & Simple Rejection & 56.9 & 74.2 & 35.2 \\
          & Score-based      & 55.7 & 73.5 & 35.5 \\
          & Best Model       & 56.4 & 74.0 & 35.3 \\
\bottomrule
\end{tabular}%
}
\endgroup
\end{table}

Table~\ref{tab:eval_ct5_mr_ux} shows that the three-version NvRec achieves high TAR across filter and decision-rule combinations, with the highest TAR of \textbf{83.8\%} obtained by H-Filter + Simple Rejection.
Because NvRec rejects unreliable candidates by \textit{TailAnalyzer} and filters, improvements in TAR are accompanied by increased RR, and stricter filtering (i.e., H-Filter and RH-Filter) also tends to increase FRR, reflecting the trade-off between reliability and the risk of discarding potentially correct recommendations.


In contrast, Table~\ref{tab:eval_results_all_five} shows that the five-version NvRec yields consistently lower TAR than the three-version NvRec across all filtering strategies and decision rules. The best TAR achieved by the configuration with R-Filter + Simple Rejection is only 67.1\%, about 15\% lower than the corresponding three-version NvRec. The result suggests that adding more versions does not always improve the reliability of recommendations with respect to TAR. Since TAR is counted over the outputs after tail rejection and filtering, rather than over all the inputs, the effectiveness of the filters significantly affects the results. In the five-version NvRec, despite greater diversity than in the three-version case, RR is relatively lower by filters while FRR remains in the same range, resulting in more incorrect final outputs. We will further show in the next section that the five-version NvRec without filtering achieves a higher TAR, highlighting the ineffectiveness of filters in this case.

\vspace{2ex}
\noindent\begin{tikzpicture}
\node [draw,
rounded corners=5pt,
fill=gray!10,
inner sep=8pt,
align=justify,
text width=0.95\columnwidth] {
\textbf{Answer to RQ2:} NvRec improves recommendation reliability by combining multiple models with filtering and output decision rules.
Across three-model configurations, H-Filter with Simple Rejection consistently achieves the highest TAR within each combination, and the best-performing three-version configuration is CodeT5\&MulaRec\&UniXcoder. 
However, five versions NvRec degrades TAR across all filter and decision-rule settings due to weaker agreement among models.
};
\end{tikzpicture}

\subsubsection{Impact of Output Decision Rules}
\label{subsubsec:impact_of_output_decision}

\begin{table}[t]
\caption{Comparison of Output Decision Rules without Filtering (Three Models)}
\label{tab:output_decision_methods}
\centering
\begingroup
\resizebox{0.75\columnwidth}{!}{%
\begin{tabular}{lrrr}
\toprule
\textbf{Decision Rule} &
\textbf{TAR (\%)} &
\textbf{RR (\%)} &
\textbf{FRR (\%)} \\
\midrule
Simple Rejection & $\mathbf{82.7}$ & 68.1 & 23.2 \\
Score-based      & 78.3 & 65.9 & 24.0 \\
Best Model       & 80.7 & 67.2 & 23.5 \\
\bottomrule
\end{tabular}
}
\endgroup
\end{table}

\begin{table}[t]
\caption{Comparison of Output Decision Rules without Filtering (Five Models)}
\label{tab:output_decision_methods_five}
\centering
\begingroup
\resizebox{0.75\columnwidth}{!}{%
\begin{tabular}{lrrr}
\toprule
\textbf{Decision Rule} &
\textbf{TAR (\%)} &
\textbf{RR (\%)} &
\textbf{FRR (\%)} \\
\midrule
Simple Rejection & $\mathbf{83.1}$ & 69.0 & 21.9 \\
Score-based      & 75.2 & 65.2 & 23.2 \\
Best Model       & 81.0 & 68.0 & 22.2 \\
\bottomrule
\end{tabular}
}
\endgroup
\end{table}

We evaluate the impact of output decision rules on recommendation reliability by comparing three strategies under settings without filtering.

We first evaluate the results for the three-version configuration shown in Table~\ref{tab:output_decision_methods}.
Among the three decision rules, \textbf{Simple Rejection} achieves the highest TAR of \textbf{82.7\%}.
The \textbf{Best Model} strategy follows with a TAR of 80.7\%, while the \textbf{Score-based} approach yields the lowest TAR of 78.3\%.
Accordingly, the performance ranking is Simple Rejection, Best Model, and Score-based.
However, Simple Rejection also exhibits relatively high RR (68.1\%) and FRR (23.2\%), indicating that improvements in TAR come at the cost of discarding more recommendation candidates.

We next show the evaluation results of the five-version configuration in Table~\ref{tab:output_decision_methods_five}.
The relative performance ordering among decision rules remains consistent with that observed in the three-version configuration.
Simple Rejection again achieves the highest TAR (83.1\%) and the lowest FRR (21.9\%).
The Best Model strategy achieves a TAR of 81.0\%, while the Score-based approach performs substantially worse, with a TAR of 75.2\%.
The improved performance of Simple Rejection in the five-version configuration suggests that a larger number of models increases the likelihood of reaching consensus among individual recommendations via majority voting.
In contrast, the Score-based decision strategy suffers a noticeable degradation in TAR relative to the three-version configuration, suggesting that including weaker models may reduce the stability of score aggregation.
Depending on the decision rules, additional models affect TAR either positively (Simple Rejection and Best Model) or negatively (Score-based).

\vspace{2ex}
\noindent\begin{tikzpicture}
\node [draw,
rounded corners=5pt,
fill=gray!10,
inner sep=8pt,
align=justify,
text width=0.95\columnwidth] {
\textbf{Answer to RQ3:} Among the evaluated output decision rules, Simple Rejection consistently achieves the highest TAR regardless of the presence or absence of filtering, followed by the Best Model strategy. Although additional score-based processing can reduce the RR, it also decreases TAR, yielding limited overall benefits.
};
\end{tikzpicture}

\subsubsection{Effectiveness of Filtering}
\label{subsubsec:filter_effectiveness}
To evaluate the effectiveness of filtering in the three-version NvRec, we first compare the results presented in Table~\ref{tab:eval_ct5_mr_ux}.
The results indicate a clear tendency that applying stricter filters leads to higher TAR.
In particular, under the Simple Rejection rule, TAR increases from 82.9\% with R-Filter to 83.8\% with H-Filter, and remains comparably high at 83.7\% with RH-Filter.
Moreover, across all output decision rules, H-Filter consistently achieves the highest TAR, with values ranging from 81.5\% to 83.8\%. At the same time, the improvement in TAR is accompanied by an increase in the FRR.
With R-Filter, the FRR ranges from 28.5\% to 29.2\%, whereas it increases to 32.2\%--32.5\% with H-Filter and further to 33.8\%--34.1\% with RH-Filter.
These results demonstrate that, in the three-version configuration, stricter filtering introduces a trade-off between higher TAR and increased false rejection.

We next examine the effect of filtering in the five-version NvRec using the results shown in Table~\ref{tab:eval_results_all_five}.
In contrast to the three-version setting, applying stricter filters in the five-version configuration degrades performance.
Under the Simple Rejection rule, TAR decreases substantially from 67.1\% with R-Filter to 60.8\% with H-Filter and further to 56.9\% with RH-Filter.
Meanwhile, the FRR increases from 27.5\% to 33.1\% and 35.2\%, respectively.
A similar trend is observed for the Score-based and Best Model decision rules, where TAR decreases to approximately 55.7\%--60.3\% under stricter filtering, while the FRR ranges from 33.2\% to 35.5\%. These results suggest that when the model ensemble includes weaker-performing models, filtering is more likely to discard candidates that could otherwise yield correct recommendations.
As a consequence, applying stricter filters can be counterproductive, ultimately reducing the quality of accepted recommendations.

\vspace{2ex}
\noindent\begin{tikzpicture}
\node [draw,
rounded corners=5pt,
fill=gray!10,
inner sep=8pt,
align=justify,
text width=0.95\columnwidth] {
\textbf{Answer to RQ4:} Comparisons among R-Filter, H-Filter, and RH-Filter indicate that stricter filtering generally tends to improve TAR. In our evaluation, H-Filter achieves the highest TAR in the three-version configuration. However, in the five-version configuration, filtering becomes counterproductive, leading to a decrease in TAR. In particular, RH-Filter, which checks both the output history and tail flags, achieves the highest RR among the evaluated filters.
};
\end{tikzpicture}

\subsection{Implications to practical usage}
The evaluation results reveal that NvRec can achieve significantly higher reliability than individual recommendation models by proactively dropping unreliable recommendations. The ability to drop unreliable recommendations is practically important as developers can reduce the risk of API misuse caused by inappropriate recommendations. Such high reliability is particularly required in high-assurance software development contexts and for novice and beginner developers. However, as observed in our empirical study, higher RRs indicate that improved reliability comes at the cost of a high rejection rate, which significantly limits the availability of the recommendation tool. Although incorrect API recommendations do not exactly match the correct answers, they may still offer hints for developing or improving the code. Thus, practical tools can also leverage rejected API recommendations to help developers with lower-confidence alerts, enabling them to carefully accept or reject the recommendations.

\section{Threats to Validity}

\subsection{Internal Validity}
Our evaluation assumes that API method sequences can be accurately extracted from model outputs.
However, errors in the extraction process could distort the evaluation results, for example, by incorrectly classifying an otherwise correct recommendation as incorrect.

To mitigate the threat, we reuse the string-parsing function used in the study by Irsan et al.~\cite{MULAREC} and unify the conversion of model outputs into lists of API method sequences.
By adopting the same evaluation infrastructure as MulaRec, we minimize threats to internal validity arising from inconsistencies in how outputs are interpreted.

\subsection{External Validity}
First, the diversity of the evaluated models is inherently limited.
In this study, we initially evaluate three-version configurations based on CodeBERT, CodeT5, and MulaRec.
Although these models have been widely used in prior work, they do not necessarily produce similar outputs for the same inputs.
To increase model diversity, we additionally include code-related models with different architectures, namely CodeT5+ and UniXcoder, in our comparative evaluation.

Second, the characteristics of the dataset used for evaluation may also pose a threat to external validity.
Our experiments are conducted on a single dataset targeting Java programs, and it remains unclear whether the observed results generalize to other programming languages (e.g., Python) or to projects from different application domains.
On the other hand, as described in Section~\ref{subsec:experimental_setup}, publicly available benchmarks that provide large-scale pairs of natural language queries and source code contexts for the API method sequence recommendation task are extremely limited.
At present, the dataset released by Zhou et al.\ is effectively the only such evaluation benchmark.
Moreover, the dataset has been used in many prior studies, including Irsan et al.~\cite{MULAREC}.
From the perspective of comparability with existing work, the evaluation results reported in this study are therefore considered sufficiently valid.

As an important direction for future work, we plan to evaluate NvRec on additional datasets, including benchmarks for other programming languages and newly constructed datasets, in order to further examine the generalizability of our findings across a broader range of settings.

\section{Conclusion and Future Work}
We proposed a reliable API method sequence recommendation approach, NvRec, which leverages  
multiple ML models to suppress unreliable API recommendations entailing tail APIs.
Through extensive evaluations, the best NvRec configuration achieves a TAR of 83.8\%, a RR of 80.7\%, and a FRR of 32.2\%, demonstrating substantially higher reliability than the best individual model. 
Our results show that \textit{Tail Analyzer} solely improves the TAR of individual models. In addition, in the three-version configuration, combining models with diverse error patterns leads to higher TAR. In the five-version configuration, avoiding filtering and aggregating recommendations via simple majority voting tends to be more effective.
NvRec can be configured with different ML models, filtering rules, and output decision rules to strike a desirable balance between reliability and efficiency.

Despite these promising results, the proposed approach incurs a relatively high rejection rate. NvRec is designed to proactively eliminate recommendations deemed unreliable, which inevitably results in a non-negligible number of inputs for which no recommendation is returned. This behavior prioritizes safety and reliability, but it may reduce the opportunity for providing recommendations in practical deployment scenarios. An important direction for future work is therefore to refine filtering strategies and output decision rules in order to maintain a high TAR while reducing the rejection rate. Moreover, since the N-version inference framework is not limited to API recommendation, it has the potential to be extended to other code generation and recommendation tasks.

\bibliographystyle{ACM-Reference-Format}
\bibliography{sample-base}

@String{Computing = "Computing" }

@String{Computer = "{IEEE} Computer" }

@inproceedings{DeepAPI,
  author    = {Gu, Xiaodong and Zhang, Hongyu and Zhang, Dongmei and Kim, Sunghun},
  title     = {Deep API Learning},
  booktitle = {Proceedings of the 24th ACM SIGSOFT International Symposium on Foundations of Software Engineering},
  series    = {FSE 2016},
  pages     = {631--642},
  year      = {2016},
  doi       = {10.1145/2950290.2950334},
}

@inproceedings{TheDevil,
  author    = {Zhou, Xin and Kim, Kyungmin and Xu, Bowen and Liu, Junjie and Han, Dong and Lo, David},
  title     = {The Devil is in the Tails: How Long-Tailed Code Distributions Impact Large Language Models},
  booktitle = {Proceedings of the 38th IEEE/ACM International Conference on Automated Software Engineering},
  series    = {ASE 2023},
  pages     = {40--52},
  year      = {2023},
  doi       = {10.1109/ASE56229.2023.00157},
}

@article{LearningAPIisDifficult,
    author = {Scaffidi, Christopher},
    title = {Why are APIs difficult to learn and use?},
    publisher = {Association for Computing Machinery},
    address = {New York, NY, USA},
    volume = {12},
    number = {4},
    issn = {1528-4972},
    doi = {10.1145/1144359.1144363},
    journal = {XRDS},
    pages = {4},
    year = {2006},
    numpages = {1}
}

@INPROCEEDINGS{Identifier-based,
    author={L. Heinemann and V. Bauer and M. Herrmannsdoerfer and B. Hummel},
    booktitle={the 16th European Conference on Software Maintenance and Reengineering}, 
    title={Identifier-Based Context-Dependent API Method Recommendation}, 
    pages={31-40},
    year={2012},
    doi={10.1109/CSMR.2012.14}
}

@inproceedings{MULAREC,
  author    = {Irsan, Ilham Cahya and Zhang, Tianyi and Thung, Ferdian and Kim, Kwangkeun and Lo, David},
  title     = {Multi-Modal API Recommendation},
  booktitle = {Proceedings of the IEEE International Conference on Software Analysis, Evolution and Reengineering},
  series    = {SANER 2023},
  pages     = {272--283},
  year      = {2023},
  doi       = {10.1109/SANER56733.2023.00034},
}

@inproceedings{CodeBERT,
  author    = {Feng, Zhangyin and Guo, Daya and Tang, Duyu and Duan, Nan and Feng, Xiaocheng and Gong, Ming and Shou, Linjun and Qin, Bing and Liu, Ting and Jiang, Daxin and Zhou, Ming},
  title     = {CodeBERT: A Pre-Trained Model for Programming and Natural Languages},
  booktitle = {Findings of the Association for Computational Linguistics: EMNLP 2020},
  pages     = {1536--1547},
  year      = {2020},
  doi       = {10.18653/v1/2020.findings-emnlp.139},
}

@inproceedings{CodeT5,
  author    = {Wang, Yue and Wang, Weishi and Joty, Shafiq and Hoi, Steven C. H.},
  title     = {CodeT5: Identifier-Aware Unified Pre-trained Encoder-Decoder Models for Code Understanding and Generation},
  booktitle = {Proceedings of the 2021 Conference on Empirical Methods in Natural Language Processing},
  pages     = {8696--8708},
  year      = {2021},
  doi       = {10.18653/v1/2021.emnlp-main.685},
}

@misc{UniXcoder,
      title={UniXcoder: Unified Cross-Modal Pre-training for Code Representation}, 
      author={Daya Guo and Shuai Lu and Nan Duan and Yanlin Wang and Ming Zhou and Jian Yin},
      year={2022},
      eprint={2203.03850},
      archivePrefix={arXiv},
      primaryClass={cs.CL},
}

@misc{CodeT5P,
      title={CodeT5+: Open Code Large Language Models for Code Understanding and Generation}, 
      author={Yue Wang and Hung Le and Akhilesh Deepak Gotmare and Nghi D. Q. Bui and Junnan Li and Steven C. H. Hoi},
      year={2023},
      eprint={2305.07922},
      archivePrefix={arXiv},
      primaryClass={cs.CL},
      url={https://arxiv.org/abs/2305.07922}, 
}

@article{DDASR,
  author    = {Nan, Siyu and Wang, Jian and Zhang, Neng and Li, Duantengchuan and Li, Bing},
  title     = {DDASR: Deep Diverse API Sequence Recommendation},
  journal   = {ACM Transactions on Software Engineering and Methodology},
  articleno = {162},
  numpages  = {39},
  year      = {2025},
}

@inproceedings{50K-C,
    author={Martins, Pedro and Achar, Rohan and Lopes, Cristina V.},
    booktitle={2018 IEEE/ACM 15th International Conference on Mining Software Repositories (MSR)}, 
    title={50K-C: A Dataset of Compilable, and Compiled, Java Projects}, 
    pages={1--5},
    year={2018},
}

@misc{AgentCoder,
    title={AgentCoder: Multi-Agent-based Code Generation with Iterative Testing and Optimisation}, 
    author={Dong Huang and Jie M. Zhang and Michael Luck and Qingwen Bu and Yuhao Qing and Heming Cui},
    year={2024},
    eprint={2312.13010},
    archivePrefix={arXiv},
    primaryClass={cs.CL},
}

@article{MetaGPT,
    title={MetaGPT: Meta Programming for A Multi-Agent Collaborative Framework}, 
    author={Sirui Hong and Mingchen Zhuge and Jiaqi Chen and Xiawu Zheng and Yuheng Cheng and Ceyao Zhang and Jinlin Wang and Zili Wang and Steven Ka Shing Yau and Zijuan Lin and Liyang Zhou and Chenyu Ran and Lingfeng Xiao and Chenglin Wu and Jürgen Schmidhuber},
    year={2024},
    journal={arXiv preprint arXiv:2308.00352},
    primaryClass={cs.AI},
}

@article{CAPIR,
    title={Compositional API Recommendation for Library-Oriented Code Generation},
    author={Ma, Zexiong and An, Shengnan and Xie, Bing and Lin, Zeqi},
    journal={In the IEEE/ACM 32nd International Conference on Program Comprehension (ICPC)},
    pages={87--98},
    year={2024},
}

@inproceedings{Toolformer,
    title={Toolformer: Language Models Can Teach Themselves to Use Tools},
    author={Timo Schick and Jane Dwivedi-Yu and Roberto Dessi and Roberta Raileanu and Maria Lomeli and Eric Hambro and Luke Zettlemoyer and Nicola Cancedda and Thomas Scialom},
    booktitle={Thirty-seventh Conference on Neural Information Processing Systems},
    year={2023},
}

@inproceedings{Gorilla,
    author = {Patil, Shishir G. and Zhang, Tianjun and Wang, Xin and Gonzalez, Joseph E.},
    booktitle = {Advances in Neural Information Processing Systems},
    editor = {A. Globerson and L. Mackey and D. Belgrave and A. Fan and U. Paquet and J. Tomczak and C. Zhang},
    pages = {126544--126565},
    title = {Gorilla: Large Language Model Connected with Massive APIs},
    volume = {37},
    year = {2024}
}

@INPROCEEDINGS{N-VERSION,
    author={Liming Chen and Avizienis, A.},
    booktitle={Twenty-Fifth International Symposium on Fault-Tolerant Computing, 1995, ' Highlights from Twenty-Five Years'.}, 
    title={N-VERSION PROGRAMMINC: A FAULT-TOLERANCE APPROACH TO RELlABlLlTY OF SOFTWARE OPERATlON}, 
    volume={},
    number={},
    pages={113-},
    year={1995},
    doi={10.1109/FTCSH.1995.532621}}

@inproceedings{APIChange,
    author = {Linares-V\'{a}squez, Mario and Bavota, Gabriele and Bernal-C\'{a}rdenas, Carlos and Di Penta, Massimiliano and Oliveto, Rocco and Poshyvanyk, Denys},
    title = {API change and fault proneness: a threat to the success of Android apps},
    booktitle = {Proceedings of the 2013 9th Joint Meeting on Foundations of Software Engineering},
    pages = {477--487},
    year = {2013},
    numpages = {11},
    series = {ESEC/FSE 2013}
}

@INPROCEEDINGS{ImprovingAPIDoc,
    author={Dekel, Uri and Herbsleb, James D.},
    booktitle={2009 IEEE 31st International Conference on Software Engineering}, 
    title={Improving API documentation usability with knowledge pushing}, 
    volume={},
    number={},
    pages={320--330},
    year={2009},
}

@misc{JavaAPI,
  title        = {{Java Platform, Standard Edition Java Development Kit Version 17 API Specification}},
  author       = { },
  organization = {{Oracle Corporation}},
  address      = {Austin, TX},
  year         = {2021},
  month        = {September},
  note         = {URL: \url{https://docs.oracle.com/en/java/javase/17/docs/api/}},
}

@ARTICLE{WhatAPIisHard,
  author={Robillard, Martin P.},
  journal={IEEE Software}, 
  title={What Makes APIs Hard to Learn? Answers from Developers}, 
  year={2009},
  volume={26},
  number={6},
  pages={27--34},
  doi={10.1109/MS.2009.193}
}

@article{FieldStudyAPI,
author = {Robillard, Martin P. and Deline, Robert},
title = {A field study of API learning obstacles},
year = {2011},
journal = {Empirical Softw. Engg.},
volume = {16},
number = {6},
pages = {703--732},
numpages = {30},
}

@article{HowToUseAPIDoc,
author = {Meng, Michael and Steinhardt, Stephanie and Schubert, Andreas},
title = {How developers use API documentation: an observation study},
year = {2019},
volume = {7},
number = {2},
journal = {Commun. Des. Q. Rev},
pages = {40--49},
numpages = {10},
}

@INPROCEEDINGS{API-MisuseBug,
  author={Gu, Zuxing and Wu, Jiecheng and Liu, Jiaxiang and Zhou, Min and Gu, Ming},
  booktitle={2019 IEEE 43rd Annual Computer Software and Applications Conference (COMPSAC)}, 
  title={An Empirical Study on API-Misuse Bugs in Open-Source C Programs}, 
  year={2019},
  volume={1},
  number={},
  pages={11--20},
}

@INPROCEEDINGS{API-Misuse2,
  author={Li, Xia and Jiang, Jiajun and Benton, Samuel and Xiong, Yingfei and Zhang, Lingming},
  booktitle={2021 14th IEEE Conference on Software Testing, Verification and Validation (ICST)}, 
  title={A Large-scale Study on API Misuses in the Wild}, 
  year={2021},
  volume={},
  number={},
  pages={241--252},
}

@article{machida2023using,
  title={Using diversities to model the reliability of two-version machine learning systems},
  author={Machida, Fumio},
  journal={IEEE Transactions on Emerging Topics in Computing},
  volume={12},
  number={3},
  pages={810--825},
  year={2023},
  publisher={IEEE}
}

@article{avizienis1985n,
  title={The N-version approach to fault-tolerant software},
  author={Avizienis, Algirdas},
  journal={IEEE Transactions on software engineering},
  number={12},
  pages={1491--1501},
  year={1985},
  publisher={IEEE}
}

\end{document}